# Frequency conversion of vortex states by chiral forward Brillouin scattering in twisted photonic crystal fibre

Xinglin Zeng[1*], Philip St.J. Russell[1] and Birgit Stiller[1,2*]

[1]Max Planck Institute for the Science of Light, Staudtstr. 2, 91058 Erlangen, Germany
[2]Department of Physics, Friedrich-Alexander Universität, Staudtstr. 7, 91058 Erlangen, Germany

[*]Corresponding author: xinglin.zeng@mpl.mpg.de, birgit.stiller@mpl.mpg.de

**Optical vortex states—higher optical modes with helical phase progression and carrying orbital angular momentum—have been explored to increase the flexibility and capacity of optical fibres employed for example in mode-division-multiplexing, optical trapping and multimode imaging. A common requirement in such systems is high fidelity transfer of signals between different frequency bands and modes, which for vortex modes is not so straightforward. Here we report intervortex conversion between backward-propagating circularly polarised vortex modes at one wavelength, using chiral flexural phonons excited by chiral forward stimulated Brillouin scattering at a different wavelength. The experiment is carried out using chiral photonic crystal fibre, which robustly preserves circular polarisation states. The chiral acoustic wave, which has the geometry of a spinning single-spiral corkscrew, provides the orbital angular momentum necessary to conserve angular momentum between the coupled optical vortex modes. The results open up new opportunities for inter-band optical frequency conversion and the manipulation of vortex states in both classical and quantum regimes.**

## 1. Introduction

Transfer of electromagnetic signals between different wavelength bands is crucial for next-generation classical and quantum information technologies[1–4]. It holds the key to viable implementation of optical networks, which often require interfacing of several subsystems operating in widely different spectral regions. Frequency conversion can be achieved using second or third order nonlinear processes[5–8], stimulated Raman scattering[9,10], or phonon-photon interactions[4,11–13], which depending on the system design can be used to shift the frequency of an optical signal by up to 10s of GHz[14]. This can be realised for example using electro-optomechanical[15] or piezo-optomechanical effects[16], or stimulated Brillouin-Mandelstam scattering (SBS)[17–19]. The first two approaches require complex setups to efficiently drive the phonons, whereas SBS is often easier to implement in practice.

Optical information can be encoded in variables such as amplitude, phase, polarisation states and spatial mode, the last two of which usually require the use of special optical fibres. Over the last decade, there has been a drive to develop sensing, microscopy and communications systems based on different types of multimode optical fibres[20–22]. The modes of such fibres can be expanded as spatial modes, carrying no orbital angular momentum, or alternatively as vortex modes. Although vortex modes in conventional step-index fibres have been used in optical communications[23], light trapping[24] and quantum information processing[25], their utility is restricted by poor preservation of circular polarisation state. The recent advent of chiral photonic crystal fibre (PCF), which can robustly preserve azimuthal order and circular polarisation state[26], have overcome this limitation and makes themselves particularly interesting for studying phonon-photon interaction of structured optical waves[27–29].

Here, we report frequency conversion by SBS between forward-propagating vortex modes in chiral PCF, enabling investigation of intervortex photon-phonon scattering in the presence of chirality. The beat-note between pump and Stokes modes drives the creation of phonons that are themselves vortices or "chiral flexural waves" (CFWs). One example is the superposition of two orthogonal π/2-out-of-phase flexural waves, which produces an acoustic strain pattern that resembles a rotating single-spiral cork-screw. CFWs created by forward stimulated Brillouin scattering (FSBS) not only provide the orbital angular momentum necessary for frequency conversion between optical vortex modes, but can also be used for nonreciprocal phase-matched intermodal conversion at a widely separated wavelength, potentially enabling transfer of signals between wavelength bands via the coherent phonon population.

Note: throughout the paper quantities related to the pump and Stokes signals are subscripted *P* and *S*, and we adopt the convention that the pump always has a higher frequency than the Stokes.

## 2. Chiral FSBS between helical Bloch modes

Chiral PCFs with *N*-fold rotational symmetry support helical Bloch modes (hBMs)[30,31], whose *m*-th order azimuthal harmonic carries an optical vortex of order $\ell_A^{(m)} = \ell_0^{(m)} + Nm$ where $\ell_A^{(m)}$ is the number of complete periods of phase progression around the azimuth for fields expressed in cylindrical components and $\ell_A^{(0)}$ is the principal azimuthal order. Note that $\ell_A^{(0)}$ is always an integer and is robustly conserved. In chiral PCF it is found, both experimentally and by numerical modelling, that the fields are almost perfectly circularly polarized, under which circumstances the topological charge $\ell^{(m)}$ (the number of on-axis discontinuities for fields evaluated in Cartesian coordinates) is linked to the azimuthal order by $\ell^{(m)} = \ell_A^{(m)} - s$, where $s$ is the spin ($s = +1$ for left-circular (LCP) and $-1$ for right-circular (RCP) polarisation state). Here, we use the shorthand $[\ell, s]$ to denote the parameters of a hBM, where for ease of notation the principal topological order is defined as $\ell^{(0)} \equiv \ell$. In chiral PCF, hBMs with equal and opposite values of $\ell$ are generally non-degenerate in index, i.e., topologically birefringent, while modes with opposite spin but the same value of $\ell$ typically display weak circular birefringence[26].

### 2.1 Optoacoustic coupling between hBMs

In the Cartesian laboratory frame, the transverse field of a circularly polarized hBM can be written as[32]:

$$\begin{aligned}\mathbf{e}_k &= a_k(z)\mathbf{e}_{Bk} = a_k(z)\begin{pmatrix} e_{Bkx} \\ e_{Bky} \end{pmatrix} \\ &= a_k(z)\begin{pmatrix} 1 \\ s_k i \end{pmatrix}\frac{1}{\sqrt{2}}B_k(\rho, \phi + \alpha z)e^{i(\beta_k z + \ell_k \phi - \omega_k t)}\end{aligned} \tag{1}$$

where $a_k(z)$ is a slowly-varying amplitude, $B_k(\rho, \phi + \alpha z)$ is the azimuthally periodic field distribution of the *k*-th hBM ($k = P, S$), $(\rho, \phi, z)$ are cylindrical coordinates, $\alpha$ is the twist rate (rad/m), $\omega_k$ the angular frequency, $s_k = \pm 1$ the spin, and $\ell_k$ the topological charge. Note that the hBM field pattern $B_k$ rotates with $z$ at a rate $\partial\phi/\partial z = -\alpha$, locking it to the structure of the chiral fibre. $B_k$ can be expressed as the sum over azimuthal harmonics, with the *m*-th harmonic having propagation constant $\beta_k^{(m)} = \beta_k + mN\alpha$ and topological charge $\ell_k^{(m)} = \ell_k + mN$ (to simplify the notation we have set $\beta_k^{(0)} = \beta_k$ and $\ell_k^{(0)} = \ell_k$). For more details see Supplementary Materials Sec.1.1.

In the absence of any perturbation ($a_P$ and $a_S$ constant), the expression in Eq. (1) is itself a solution of Maxwell's equations in the chiral PCF. In the presence of a dielectric constant perturbation $\varepsilon_{ph}$ in the form:

$$[\varepsilon_{ph}] = \begin{bmatrix} \cos\zeta & -\sin\zeta \\ \sin\zeta & \cos\zeta \end{bmatrix} \begin{bmatrix} \varepsilon_1(\rho,\phi)/2 & 0 \\ 0 & -\varepsilon_1(\rho,\phi)/2 \end{bmatrix} \quad (2)$$

where $\zeta = Kz + \ell_{ph}\phi - \Omega t$, $\Omega = \omega_{\rm P} - \omega_{\rm S}$, $K$ is the acoustic wavevector at frequency $\Omega$, $\ell_{ph}$ is the acoustic topological charge and $\varepsilon_1$ is an induced anisotropic change in dielectric constant that yields linear birefringence (needed to couple LCP and RCP light), the coupling rate from mode $k$ to mode $l$ will be proportional to $\langle \mathbf{e}_{Bl}^{\dagger}|\varepsilon_{ph}|\mathbf{e}_{Bk}\rangle$ where † denotes the Hermitian conjugate. The rotation matrix in Eq. (2) ensures that the pattern of linear birefringence rotates with position, azimuthal angle and time, resembling a rotating single-spiral corkscrew. A chiral flexural wave (CFW) with this geometry can be produced by a superposition of two, π/2 out-of-phase, orthogonal flexural waves differing in frequency by Ω (see next section).

Evaluating $\langle \mathbf{e}_{Bl}^{\dagger}|\varepsilon_{ph}|\mathbf{e}_{Bk}\rangle$ in a coupled-mode description, assuming slowly-varying power-normalized amplitudes, a separately excited CFW (i.e., no FSBS), conservation of topological charge, i.e., $\ell_{ph} = \ell_S - \ell_P$, and collecting terms with the slowest rates of phase progression, we obtain (see Supplementary Materials Sec.1.2 for details):

$$\frac{\partial a_P}{\partial z} + \frac{\alpha_P}{2} a_P = i\kappa_P a_S e^{i\vartheta z}, \qquad \frac{\partial a_S}{\partial z} + \frac{\alpha_S}{2} a_S = i\kappa_S a_P e^{-i\vartheta z}. \quad (3)$$

where the quantity $\vartheta = \beta_S(\omega_S) - \beta_P(\omega_P) - K(\Omega)$ is the dephasing rate, $n_k$ is the phase index of mode $k$, and $\kappa_k = \omega_k Q/c$, where $Q$ is a dimensionless parameter that is proportional to the integral of $\langle \mathbf{e}_{Bl}^{\dagger}|\varepsilon_{ph}|\mathbf{e}_{Bk}\rangle$ over the fibre cross-section.

## 2.2 Forward stimulated Brillouin scattering between hBMs

The coupled FSBS equations for complex slowly-varying field amplitudes $a_i$ that scale with the square-root of the modal power ($p_i = |a_i|^2$) can be written in the form:

$$\frac{\partial a_P}{\partial z} + \frac{\alpha_P}{2} a_P = i\frac{\omega_P}{c}\sqrt{Q_B}\, a_S a_{ph}, \qquad \frac{\partial a_S}{\partial z} + \frac{\alpha_S}{2} a_S = i\frac{\omega_S}{c}\sqrt{Q_B}\, a_P a_{ph}^*,$$
$$\frac{\partial a_{ph}}{\partial z} + \left(\frac{\alpha_{ph}}{2} + i\vartheta\right) a_{ph} = i\frac{\omega_{ph}}{c}\sqrt{Q_B}\, a_S^* a_P \quad (4)$$

where $\vartheta = \beta_P - \beta_S - \beta_{ph}$ is the dephasing parameter, $c$ is the speed of light in vacuum, and $Q_B$ is a characteristic parameter (with units $\mathrm{W}^{-1}$) that depends on the electrostrictive parameters and the optoacoustic overlap.

With the good approximation (commonly used in SBS) that $|\partial/\partial z| \ll |\alpha_{ph}/2 + i\vartheta|$, Eq. (4) can be recast in the form:

$$\frac{\partial a_P}{\partial z} + \frac{\alpha_P}{2} a_P = -\frac{g_B}{2}\left(1 - i\frac{2\vartheta}{\alpha_{ph}}\right)(a_P a_S^*) a_S \qquad \text{(a)}$$
$$\frac{\partial a_S}{\partial z} + \frac{\alpha_S}{2} a_S = \frac{g_B}{2}\left(1 + i\frac{2\vartheta}{\alpha_{ph}}\right)(a_S a_P^*) a_P \qquad \text{(b)} \quad (5)$$

where

$$g_B = \frac{g_{B0}}{1 + 4\vartheta^2/\alpha_{ph}^2} = \frac{4\varpi\omega_{ph} Q_B/(c^2\alpha_{ph})}{1 + 4\vartheta^2/\alpha_{ph}^2}\,\mathrm{m}^{-1}\mathrm{W}^{-1} \quad (6)$$

and we have used the very good approximation $\omega_P \cong \omega_S \cong (\omega_P + \omega_S)/2 = \varpi$. Comparing Eqs. (6) and (3) we see that for $\vartheta = 0$:

$$Q = \frac{4\omega_{ph} Q_B}{c\alpha_{ph}} |a_P a_S^*| \tag{7}$$

which will be useful later in analysing the backward pump-to-Stokes coupling, which is essentially a linear scattering process. We also need to check the phase of the CFW at every point along the fibre. Setting $a_i = q_i(z)e^{i\psi_i(z)}$ in Eq. (5) where $q_i$ and $\psi_i$ are real-valued and extracting the imaginary parts, we obtain:

$$\frac{\partial \psi_P}{\partial z} = \vartheta \frac{g_B}{\alpha_{ph}} q_S^2\,, \qquad \frac{\partial \psi_S}{\partial z} = \vartheta \frac{g_B}{\alpha_{ph}} q_P^2, \tag{8}$$

which shows that in general the relative phase of pump and Stokes varies in a non-trivial manner along the fibre. In the experiment, however, the dephasing is zero in the writing process, so that the pump and Stokes phases are constant.

Multiplying Eq. (5a) by $a_P^*$ and Eq. (5b) by $a_S^*$ and adding each equation to its complex conjugate, we obtain the coupled power equations:

$$\frac{\partial p_P}{\partial z} + \alpha_P p_P = -g_B p_P p_S, \quad \frac{\partial p_S}{\partial z} + \alpha_S p_S = g_B p_P p_S. \tag{9}$$

# 3. Experiment

## 3.1 Experimental setup

The experimental setup is sketched in Fig. 1. An acoustic mode is first written (i.e., excited) by FSBS between the forward pump and Stokes modes, and then read out at a different wavelength by a backward P (or S) signal, which is down-shifted (or up-shifted) into an S (or P) signal. For the detailed descriptions on the experimental setup see Methods.

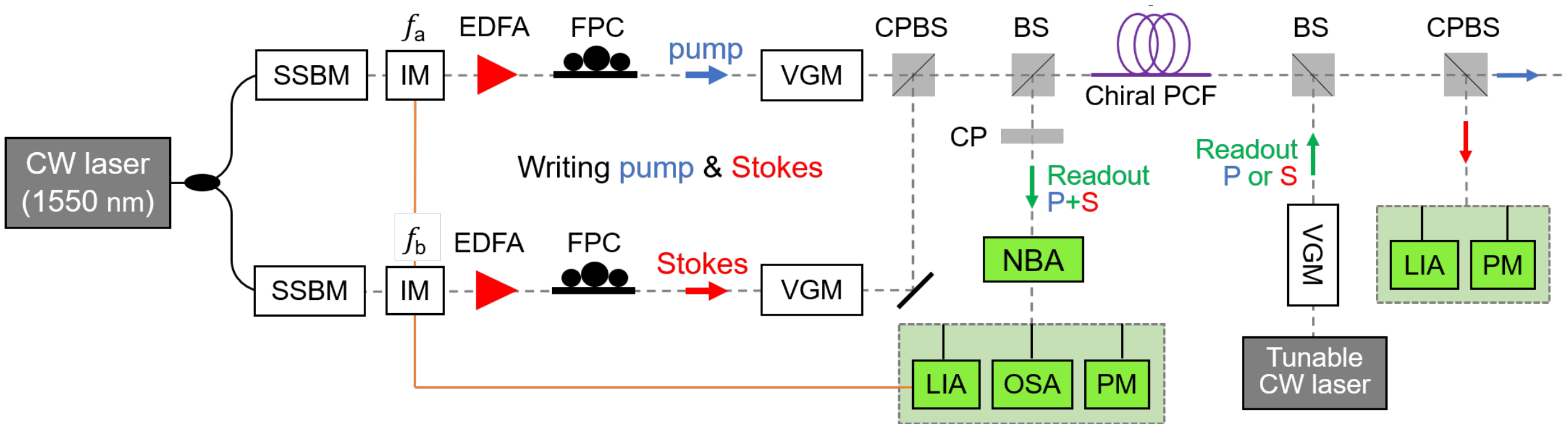

**Fig. 1.** Experimental setup for FSBS and frequency conversion between vortex modes. SSBM, single-sideband modulator; IM, intensity modulator; EDFA, erbium-doped fibre amplifier; FPC, fibre polarisation controller; VGM, vortex generation module; CPBS, circular-polarizing beam splitter; BS, beam splitter; LIA, lock-in amplifier; OSA, optical spectrum analyser, PM, power meter, NBA, near-field scanning Brillouin analyser. A coherent population of chiral phonons is written in the twisted PCF by forward-propagating pump and Stokes modes, and read out in the backward direction at a different wavelength, determined by a special phase-matching condition.

## 3.2 The chiral PCF and its modes

A three-dimensional sketch of the chiral PCF used in the experiments is shown in Fig. 2(a). The fibre has a three-fold rotationally symmetric structure with one on-axis and three satellite cores (see also scanning electron micrograph in Fig. 2(b)). The twist pitch is 5 mm, the diameter *d* of the hollow channels is 1.58 μm and the distance Λ between adjacent channels is 1.79 μm, yielding a *d*/Λ of 0.88, which results in tight confinement of both acoustic and optical fields and strong optoacoustic coupling (see Supplementary Materials Sec.2 for more details). Finite element analysis of the chiral PCF in a helicoidal frame yields the near-field intensity patterns in Fig. 2(b), showing good agreement between experiment and theory. The measurements were made after propagation along an 8 m length of the fibre, where the modulus of the Stokes parameter $|S_3|$ was higher than 0.98 at the output, showing very good preservation of circular polarisation state. The fibre loss at 1550 nm, measured by cut-back method, was 0.195 dB/m for the $\ell = 0$ modes and 0.262 dB/m for the $\ell = \pm 1$ modes.

Finite element modelling of the untwisted PCF also reveals acoustic flexural modes that are strongly confined to the core region (Fig. 2(c)). Degenerate orthogonal versions of these modes are found which, when superimposed with a π/2 phase difference, produce a wave with the topology of a rotating single-spiral corkscrew, as shown in the Fig. 2(d) (see Supplementary videos).

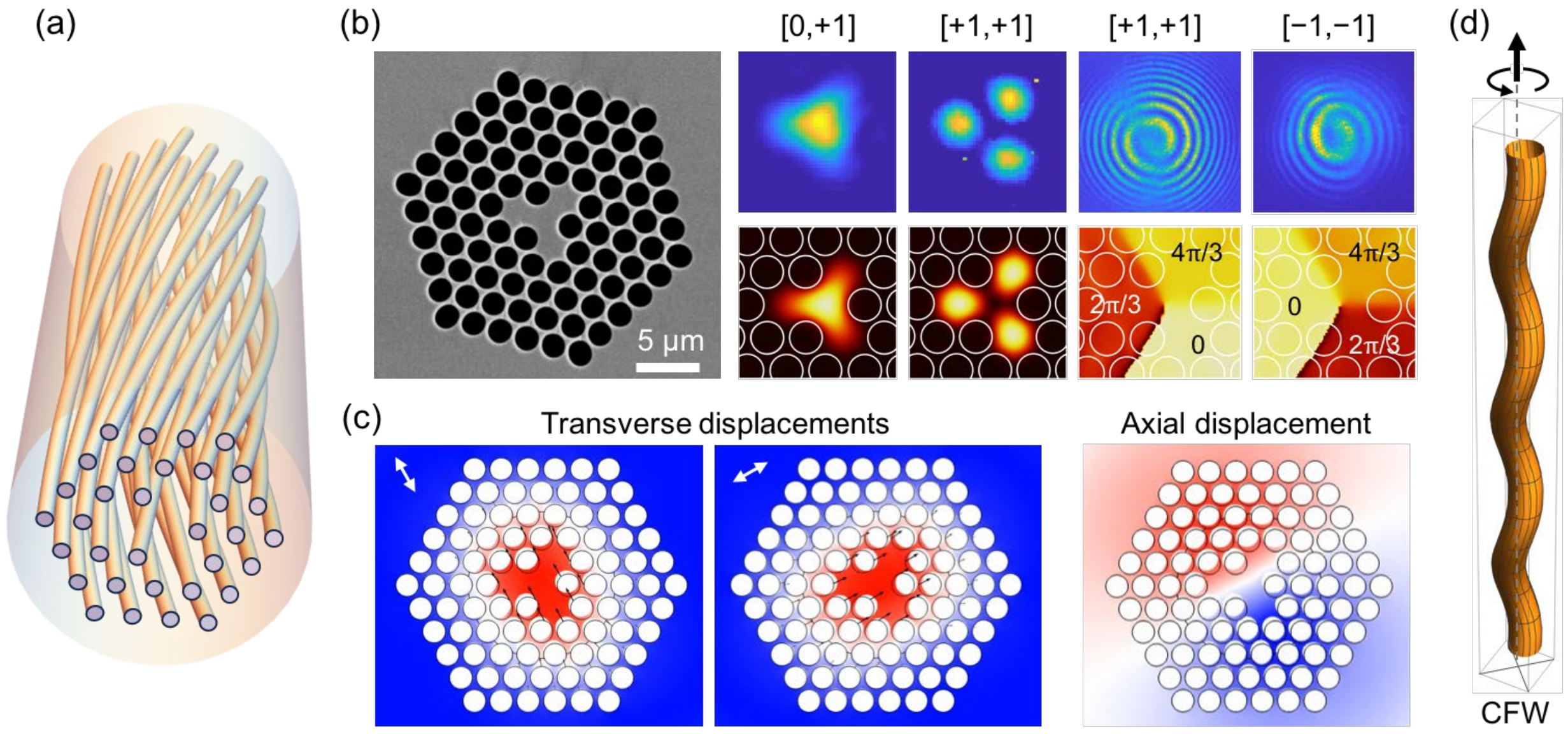

**Fig. 2.** (a) Sketch of a short length of the three-fold rotationally symmetric PCF. The small circles mark the positions of the hollow channels, embedded in fused silica. (b) Left to right: Scanning electron micrograph of the PCF structure. Upper: Measured near-field intensity patterns of the $[\ell, s] = [0, +1], [+1, +1]$ hBMs, and the spiral interference fringes that form between the modal far-fields and a divergent Gaussian beam. The intensity patterns for the $[-1, -1]$ and $[+1, +1]$ modes are very similar. Lower: Numerical simulations of the modal intensity distribution and the azimuthal phase variation of the modes. (c) Left two panels: Numerical simulations of the transverse displacements of two orthogonal degenerate flexural modes at ~96 MHz. When superimposed with a π/2 phase shift these modes generate a CFW. Right-hand panel: axial displacement of the acoustic mode, confirming the presence of a flexural wave. (d) Sketch of a CFW formed by the superposition of two orthogonal π/2-out-of-phase flexural waves. When the waves have different frequencies (as in FSBS), the shape rotates in time.

Figure 3(a) shows the $\omega$-$\beta$ diagram for the six hBMs considered in this paper, calculated by FEM and plotted over the spectral regions of interest. There is strong topological birefringence between modes with different topological charge (the calculated refractive index differences at 1550 nm are $n_{[0,+1]} - n_{[+1,+1]} = 0.034$ and $n_{[-1,-1]} - n_{[+1,+1]} = 0.00047$). In addition, the simulations confirm that modes with the same topological charge but opposite spin display weak circular birefringence, of order $8 \times 10^{-6}$. Any deviations from integer values of topological charge are caused by the polarisation state not being perfectly circular[30].

Figure 3(b) shows the $\omega$-$\beta$ diagram for three acoustic flexural modes in an untwisted fibre over the frequency range where the maximum SBS gain is seen in the experiments for $[0, +1]$ pump to $[-1, -1]$ Stokes conversion. Efficient FSBS only occurs when $\vartheta = 0$ and the optoacoustic overlap is high. Theory predicts that these conditions are fulfilled at the marked points in Fig. 3(a) and (b), corresponding to a frequency difference of ~96 MHz and an acoustic propagation constant of 0.046 rad/μm, i.e., an acoustic wavelength of 139 μm. The acoustic dispersion in the vicinity of this point is 0.0037 rad/μm per MHz, while the optical dispersion is negligible. In the chiral PCF the acoustic wavevector will differ for $\ell_{ph} = \pm 1$ by approximately $2\alpha = 4\pi/0.005 \sim 0.002$ μm$^{-1}$, which only slightly alters the phase matching condition and is neglected.

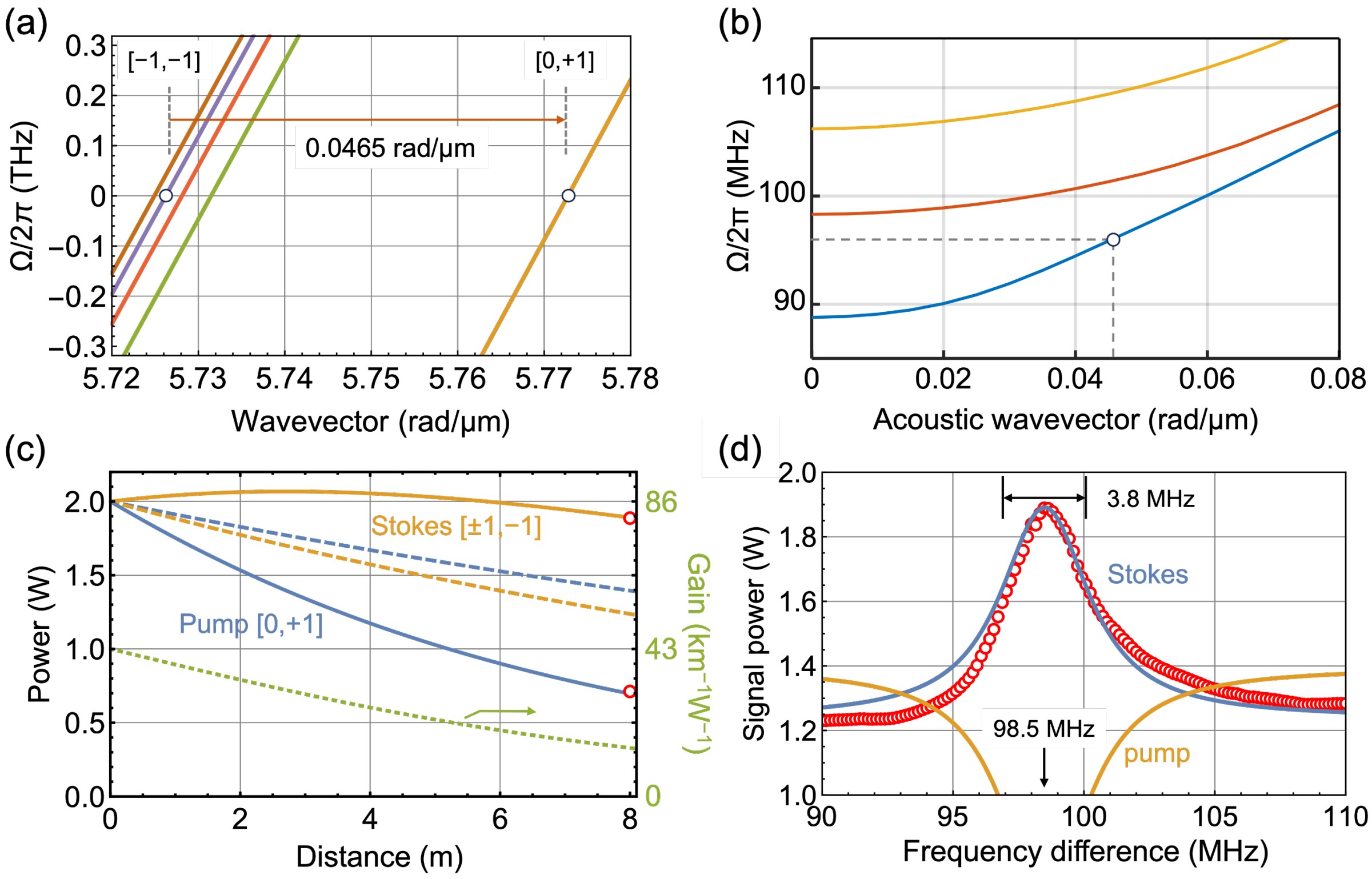

**Fig. 3.** (a) Dispersion curves for (from left to right) $[\ell, s]$ = [+1,+1], [−1,−1], [−1,+1], [+1,−1] and [0,±1] helical Bloch modes. The last two are separated by a small circular birefringence of $8\times10^{-6}$, so are indistinguishable on the scale of the plot. Zero on the frequency axis corresponds to 1550 nm. In the experiments the peak gain is seen for [0,+1] pump to [−1,−1] Stokes conversion. (b) Dispersion curves for three CFWs in an untwisted PCF, with cut-offs in the frequency range from 85 to 115 MHz, calculated by FEM. The mode with cut-off at ~89 MHz has the highest overlap with the optical modes. The open circle marks the CFW that phase-matches the [0,+1] pump and [−1,−1] Stokes modes in (a). In the vicinity of this point the acoustic dispersion is 0.0037 rad/μm per MHz. (c) Solutions of the coupled power equations for the parameters in the experiment (full curves). The red circles mark the experimental measurement. The Stokes mode experiences slight gain over the first 3 m, when there is substantial conversion from pump to Stokes. Beyond this point the FSBS conversion gradually weakens, and the signals converge to the base level exponential loss. The strength of the acoustic wave is proportional to the product of pump and Stokes, so also falls off in strength with distance. (d) The measured Stokes power as a function of frequency difference (red open circles) together with a fit to numerical solutions of Eq. (9), showing good agreement, with a full-width-half maximum of ~3.8 MHz and a peak at 98.5 MHz. The extended shoulder on the high frequency side we attribute to excitation of acoustic modes with lower overlap and higher cut-off frequencies (the upper curves in (b)).

### 3.3 Writing and read-out of chiral phonons

In the writing step, a fixed frequency pump signal is launched into the $[\ell_P, +s]$ mode, along with a frequency-tunable Stokes signal in the $[\ell_S, -s]$ mode, where $\ell_P \neq \ell_S$. Interference between these signals creates a spiralling interference pattern that, through electrostriction at the phase-match frequency, excites a CFW with topological charge of $\ell_{ph} = \ell_P - \ell_S$ (see Fig. 3(a)). Photons are converted from pump to Stokes ($\omega_P > \omega_S$) via FSBS, with the energy defect going to excite phonons. This CFW induces a spiralling 3-fold rotationally symmetric pattern of linear birefringence, which acts back on the optical fields, coupling together left and right-circularly polarized light as explained above and in the Supplementary Materials Sec.1.

### 3.3.1 Writing chiral phonons by [0,+1] to [±1,−1] scattering

In this case the CFW must supply a topological charge of $\ell_{ph} = \pm 1$, depending on which mode is the pump. The experimental setup is shown in Fig. 1 and more details are given in the Methods Section. Transmitted pump and Stokes signals were separated at a circular polarizing beam-splitter and monitored using a power meter or OSA. Peak conversion is seen at 98.5 MHz, at which point the transmitted Stokes and pump powers were 1.89 and 0.71 W respectively (in the absence of SBS, after taking account of loss, these powers would have been 1.39 and 1.23 W).

Numerically solving the coupled power equations (Eq. (9)) for $\vartheta = 0$ and adjusting the gain until the results agree with the experiment, we are able to plot the power in each mode as a function of position along the PCF (Fig. 3(c)). The red dots mark the experimental measurements, and the dashed curves the behaviour with $g_B = 0$. The green dotted curve shows the length dependence of the optoacoustic coupling, which is directly proportional to $g_{B0}|a_P a_S^*|/2$ and takes the value $g_{B0} \times 1\text{W}$ at $z = 0$. Excellent agreement with theory is obtained for $g_{B0} = 0.043\ \text{m}^{-1}\text{W}^{-1}$.

The measured half-width-half-maximum bandwidth is 1.9 MHz (Fig. 3(d)), and the lineshape fits quite well to a Lorentzian, as predicted by Eq. (6). For strong dephasing, the transmission reverts to the dashed curves in Fig. 3(c), as expected. The pronounced shoulder on the high frequency side we attribute to excitation of acoustic modes with higher frequency cut-offs (Fig. 3(b)), which have lower overlap with the optical modes. Detuning from $\vartheta = 0$ is dominated by the acoustic dispersion, which at the phase-matching frequency is 0.0037 rad/μm per MHz (Fig. 3(b)). At 1.9 MHz detuning the dephasing rate is $\vartheta = 0.0037 \times 1.9 = 0.007$ rad/μm and the gain is half of its peak value, i.e., $\alpha_{ph} = 2\vartheta = 0.014\ \mu\text{m}^{-1}$, corresponding to a 1/e acoustic power decay length of 70 μm. This in turn allows us to estimate $Q_B = 1.8 \times 10^{-5}\ \text{W}^{-1}$.

### 3.3.2 Read-out of chiral phonons by backward [±1,−1] to [0,+1] scattering

Once the CFW is excited, modal dispersion means that it can be used for phase-matched conversion in the backward direction at a shifted wavelength, as illustrated in Fig. 4(a). The black arrows represent the phonons. In the absence of group velocity dispersion, the frequency shift at which this occurs is $\Delta\Omega \approx \Omega(n_{\text{gP}} + n_{\text{gS}})/(n_{\text{gP}} - n_{\text{gS}})$[19] where $n_{gi}$ is the group index of the $i$-th mode (note that if the pump and Stokes modes are swapped, the frequency shift will change sign). In our case, however, there is significant higher order dispersion, so that this simple condition is not accurate. Numerical calculations of the error $\Delta\beta$:

$$\Delta\beta(\Delta\Omega) = \beta_{[0,+1]}(0) - \beta_{[-1,-1]}(-\Omega) - \big(\beta_{[-1,-1]}(-\Delta\Omega) - \beta_{[0,+1]}(-\Delta\Omega - \Omega)\big), \qquad (10)$$

plotted as a function of the frequency displacement $\Delta\Omega$, yield the curve in Fig. 4(b). The error falls to zero at $\Delta\Omega/2\pi = 15.5$ GHz, which is in reasonable agreement with the experimental value of 18.5 GHz. Conservation of absolute angular momentum in the forward and backward SBS processes is illustrated in Fig. 4(c). The black arrows represent the angular momentum of the phonons, which is the same in both writing and read-out. The topological charge and spin of the optical modes is however conventionally defined relative to the beam propagation direction, as in Fig. 4(a); for correct comparison with the forward process the same frame of reference must be used for both directions, so the backward signs must be reversed, as shown in the figure.

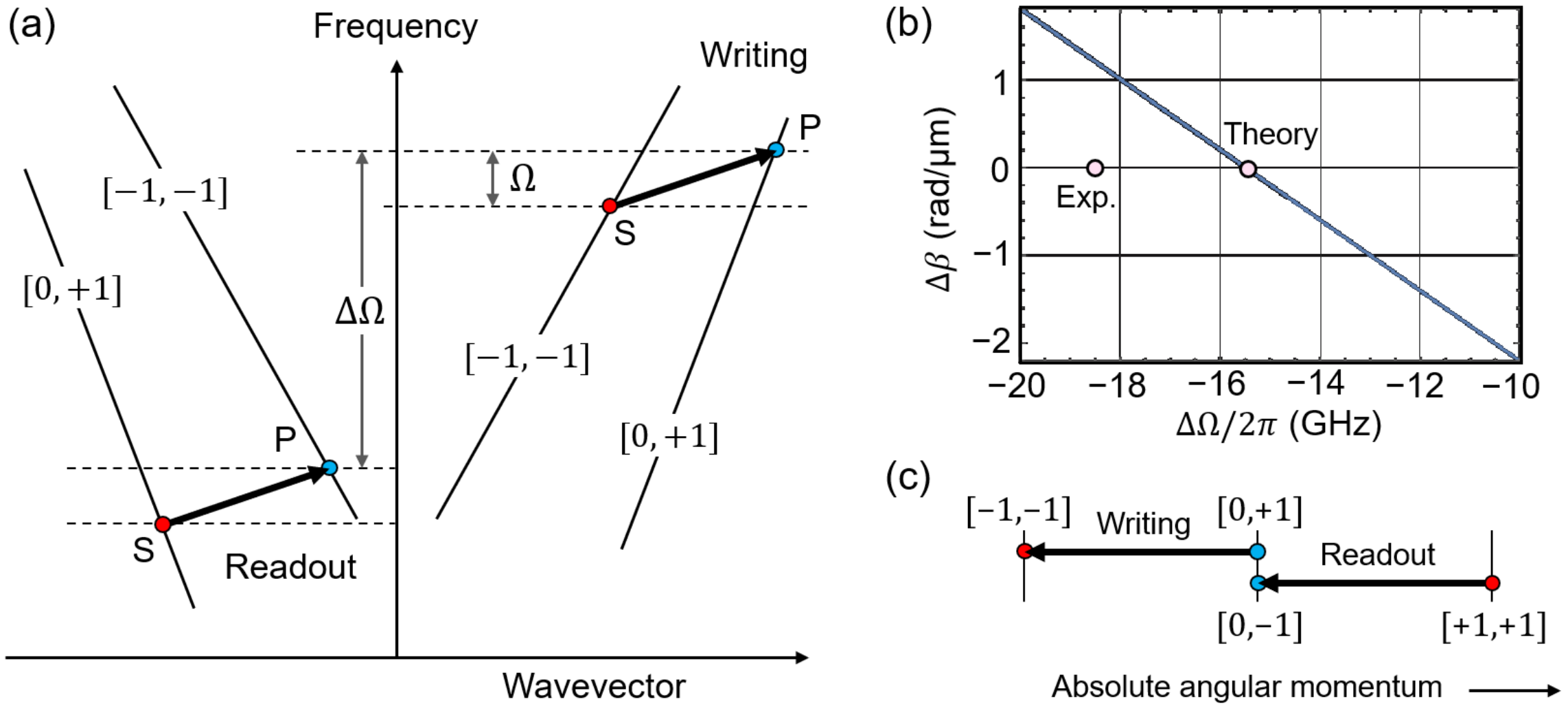

**Fig. 4.** (a) Schematic illustrating the principle of phase-matched backward read-out using a CFW excited by FSBS. The frequency displacement $\Delta\Omega$ depends on the dispersion of the two modes. Note that if the pump and Stokes modes are exchanged, the phonon propagates backwards and $\Delta\Omega$ changes sign. (b) Plot of the dephasing parameter $\Delta\beta$ for the backward process as a function of $\Delta\Omega$. The theory is based on FEM and predicts $\Delta\beta = 0$ at $\Delta\Omega/2\pi = 15.5$ GHz, in quite good agreement with the experimental value. (c) Conservation of absolute angular momentum in the forward and backward SBS processes. The black arrows represent the phonons. Note that in (a) the topological charge and spin are conventionally defined relative to the beam propagation direction; for correct comparison with the forward process, however, the same frame of reference must be used for both directions, so the backward signs must be reversed, as in the figure.

Figure 5(a) shows the read-out spectra recorded by a high resolution OSA when a $[-1,-1]$ pump signal was launched backwards into the fibre at the phase-match frequency, in the presence of strong intervortex FSBS. The horizontal scale corresponds to the frequency shift $\Delta\Omega$ of the backward pump signal relative to the forward pump signal. Strong intervortex conversion between backward pump and Stokes frequencies is observed (blue curve) when $\Delta\Omega = -18.5$ GHz. The Brillouin frequency shift $\Omega = 98.5$ MHz is identical in both directions. On the right are the modal intensity profiles of backward pump signal recorded by a CCD camera and backward Stokes signal by the NBA, confirming that the launched pump is in the $[-1,-1]$ mode and Stokes read-out in the $[0,+1]$ mode. When the pump signal is launched forwards, only very weak conversion to the backward $[0,+1]$ Stokes is seen (magenta curve), 42 dB weaker than the previous case. We attribute this signal to Rayleigh scattering and weak reflections at the fibre output face. Figure 5(b) shows the reverse process, when a $[0,+1]$ Stokes signal is launched backwards into the fibre at the phase-matching point and converted to the $[-1,-1]$ mode with a frequency up-shift of 98.5 MHz. The double peaks are caused by pixellation in the OSA. The recorded modal patterns confirm that the backward pump signal is in the $[-1,-1]$ mode.

The read-out process (subscript *r*) can be conveniently modelled using Eq. (5):

$$\frac{\partial b_{Pr}}{\partial z} + \frac{\alpha_{Pr}}{2} b_{Pr} = -\left(\frac{g_{B0}}{2} a_P a_S^*\right) b_{Sr} e^{i\vartheta_r z}, \quad \frac{\partial b_{Sr}}{\partial z} + \frac{\alpha_{Sr}}{2} b_{Sr} = \left(\frac{g_{B0}}{2} a_S a_P^*\right) b_{Pr} e^{-i\vartheta_r z} \quad (11)$$

where $b_{ir}$ are the read-out amplitudes, $g_{B0} = 0.043\ \mathrm{m^{-1}W^{-1}}$ (from Eq. (6)) and $\vartheta_r = \beta_{Sr} - \beta_{Pr} \pm K$, the sign of $K$ being chosen to minimize $|\vartheta_r|$. The distributions of $a_S$ and $a_P$ are obtained from solutions of Eq. (5) and Eq. (11) is then numerically integrated to yield the length dependence of $b_P$ and $b_S$, as shown in Fig. 5(c). From FEM calculations, the backward dephasing rate for deviations from perfect phase-matching frequency is 0.4 rad/(μm.GHz), for which solutions of Eq. (11) yield a spectral conversion efficiency that is in good agreement with experiment (Fig. 5(d)).

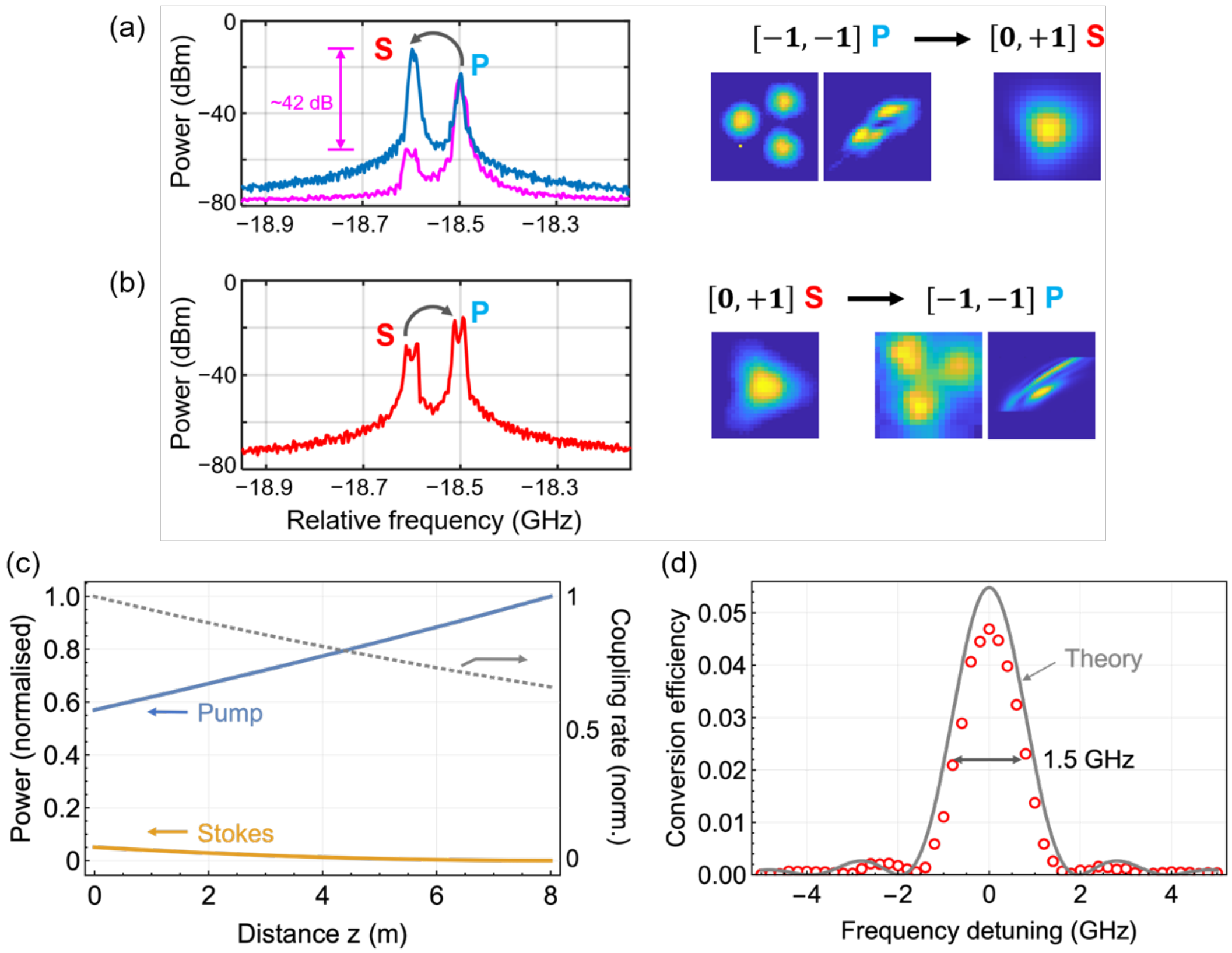

**Fig. 5.** (a) Read-out spectra recorded by OSA when a $[-1,-1]$ pump signal is launched backwards into the fibre in the vicinity of the phase-match frequency, in the presence of strong intervortex FSBS. The horizontal scale is the frequency shift of backward read-out signals over the forward writing signals. Backward FSBS is phase-matched when the backward pump frequency is shifted −18.5 GHz relative to the forward pump frequency. When the $[-1,-1]$ pump signal is launched forwards, only very weak conversion to the $[0,+1]$ Stokes is seen, 42 dB weaker than in the backward case (magenta curve), showing strong nonreciprocity. On the right are the recorded near field distributions of the backward $[-1,-1]$ pump and $[0,+1]$ Stokes. The slanted fringe pattern measured at the focus of a cylindrical lens[33] confirms the presence of an optical vortex. (b) The same as (a), but the backward intervortex conversion is from $[0,+1]$ Stokes to $[-1,-1]$ pump. The double peaks are caused by pixellation in the OSA. The recorded modal patterns confirm that the pump is in the $[-1,-1]$ mode. (c) Calculated evolution of power in backward-propagating $[-1,-1]$ pump and $[0,+1]$ Stokes read-out signals along an 8-m-long PCF in the presence of the strong acoustic wave created by intervortex FSBS (Fig. 3), with coupling constant $\kappa(z) = (g_{B0}/2)|a_S a_P^*|$, where $\kappa_0 = \kappa(0) = 0.043\ \text{m}^{-1}$. (d) Backward intervortex conversion efficiency to the Stokes with frequency detuning from perfect phase-matching, measured (red circles) and theory (grey curve) based on a dephasing rate of 0.4 (rad/μm)/GHz, calculated by FEM. Theory and experiment agree well in both peak conversion efficiency and bandwidth.

### 3.3.3 Writing and read-out with [−1,−1] and [+1,+1] modes

An expanded view of the calculated $\omega$ $\beta$ diagram for the four vortex-carrying hBMs is shown in Fig. 6(a), revealing a refractive index difference of $4.7\times10^{-4}$ between the $[+1,+1]$ and $[-1,-1]$ modes at 1550 nm, or a propagation constant difference of 0.00132 rad/μm. Figure 6(b) shows the $\omega$ $\beta$ diagram of the acoustic mode with the highest overlap with the $[+1,+1]$ and $[-1,-1]$ modes (the inset shows the $\omega$ $\beta$ curves of six other acoustic modes that have moderately high overlap with the optical modes, with cut-off frequencies in the range 1.1 to 1.6 GHz). The open circles in Figs. 6(a) and (b) mark the points where phase-matching occurs, predicting maximum gain at a Brillouin frequency shift of ~1.292 GHz for forward writing process. The mode in Fig. 6(b) comes in two degenerate orthogonal forms, which when superimposed with a frequency shift produce the rotating displacement patterns as shown in Fig. 6(c). Note that these calculations are for an untwisted PCF, see Supplementary Materials Sec.2. In the case of $[-1,-1]$ and $[+1,+1]$ writing the CFW must supply a topological charge of $\ell_{ph} = \pm 2$, which because of the three-fold symmetry is provided by the first harmonic of an $\ell_{ph} = \mp 1$ CFW. As previously reported for optical hBMs[32], higher order hBM harmonics can sometimes be stronger than those inside the first Brillouin zone, so that it is not unexpected that the SBS gain is higher for the $\ell_{ph} = \pm 2$ harmonic of the CFW.

Figure 6(d) plots the Stokes power in the writing process, measured using lock-in techniques, when the pump-Stokes frequency difference was tuned from 1.1 to 1.6 GHz. Compared to Fig. 3(d), the linewidth is much larger, suggesting that the acoustic modes in the inset in Fig. 6(b) contribute to the FSBS response, causing inhomogeneous broadening. Following the approach in Eq. (10) to calculate $\Delta\beta$ for the $[-1,-1]$ pump and $[+1,+1]$ Stokes modes, we find that backward phase-matching is predicted to occur at a wavelength of 1606.3 nm, which is shifted 6.75 THz (~54 nm) from the writing wavelength. This theoretical prediction is in excellent agreement with the experimental results in the Fig. 6(e), in which backward phase-matching occurs at 187.2 THz, corresponding to 1601.5 nm. The read-out bandwidth was measured to be 0.132 THz (1.15 nm).

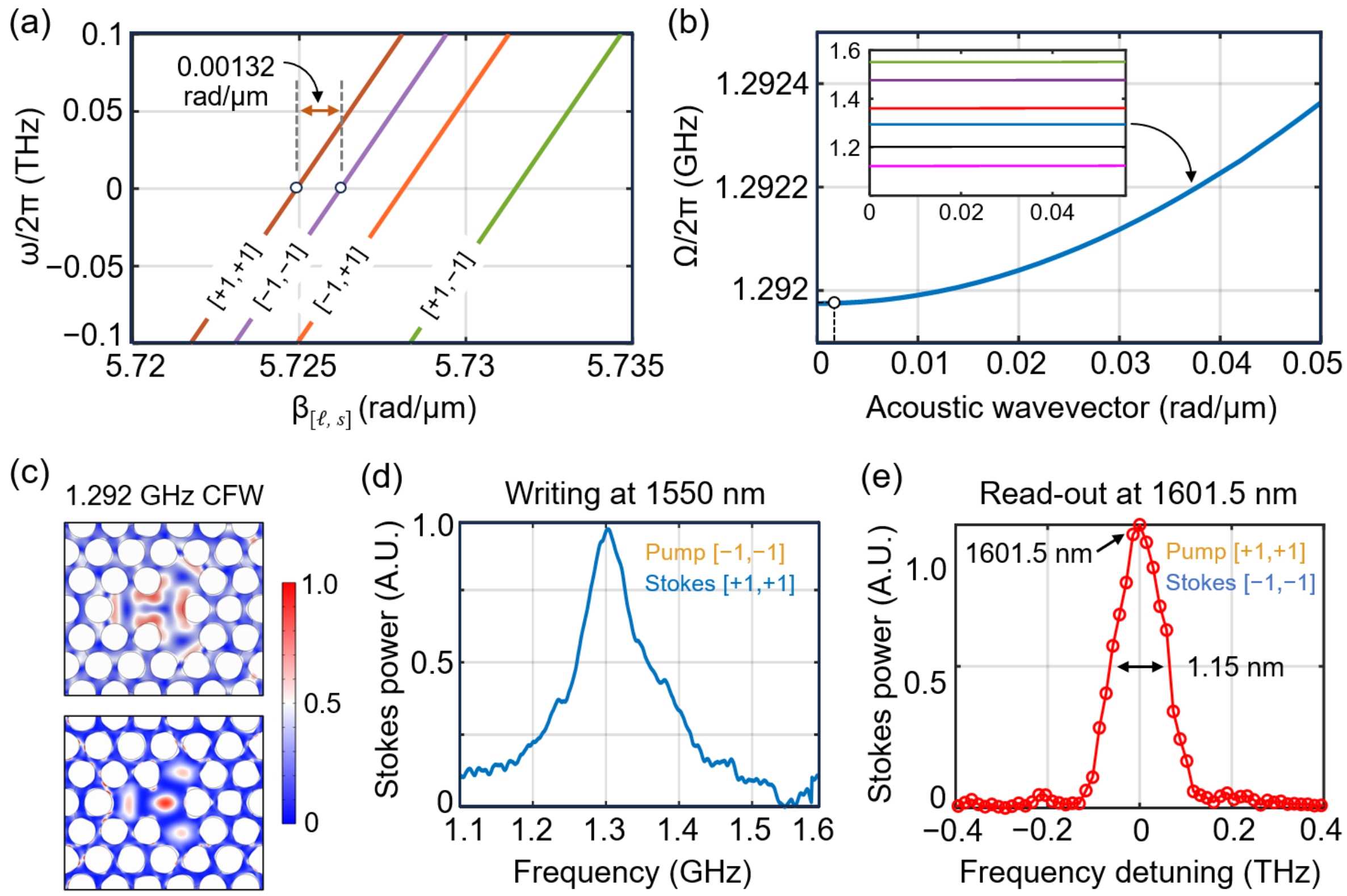

**Fig. 6.** (a) Dispersion curves for (from left to right) $[\ell, s] = [+1,+1]$, $[-1,-1]$, $[-1,+1]$, $[+1,-1]$ hBMs. Zero on the frequency axis corresponds to 1550 nm. The propagation constants of the $[-1,-1]$ and $[+1,+1]$ modes differ by 0.00132 rad/μm. (b) Calculated dispersion curves for the CFW in an untwisted PCF, with a cut-off frequency of ~1.292 GHz. The open circle marks the point at which the CFW phase-matches to the $[-1,-1]$ pump and $[+1,+1]$ Stokes modes in (a). In the vicinity of this point the acoustic dispersion is 0.136 rad/μm per MHz. The inset shows dispersion curves for neighbouring CFWs that have high overlap with the optical modes, with cut-offs frequencies in the range 1.1 to 1.6 GHz. (c) Transverse and axial displacements of the CFW at 1.292 GHz. The CFW has topological charge of $\ell_{ph} = \pm 2$, which because of the three-fold symmetry is provided by the first harmonic of an $\ell_{ph} = \mp 1$ CFW. (d) The transmitted power in the $[+1,+1]$ Stokes mode during writing, measured by the LIA, plotted as a function of frequency difference between pump and Stokes. (e) The power in the backward $[-1,-1]$ Stokes signal during read-out by a weak backward $[+1,+1]$ pump signal, measured as a function of detuning from the phase-matching frequency of 187.2 THz (~1601.5 nm), which is shifted 6.21 THz (~51.5 nm) from the writing wavelength (1550 nm).

Although a clear signal was observed at this wavelength in the experiment, confirming that it is possible to create and scatter off $\ell_{ph} = \pm 2$ flexural phonons, it was difficult to theoretically characterise the process because the response of [−1,−1] to [+1,+1] scattering were very weak.

## 4. Conclusions

Forward stimulated Brillouin scattering between optical vortex modes occurs as a result of the excitation of chiral flexural waves (CFWs) in the core region of a three-fold rotationally symmetric chiral photonic crystal fibre, which robustly preserves circularly-polarised states. Gain is seen when the pump and Stokes modes have orthogonal circular polarisation states, under which circumstances the chiral acoustic wave must induce a rotating pattern of linear birefringence, as well as providing the orbital angular momentum needed for scattering between pump and Stokes modes with differing topological charge. In the case studied in detail, the CFW has a strain field that resembles a rotating single-spiral corkscrew, and may be viewed as the superposition of two orthogonal flexural waves with different frequencies. A combination of finite-element modelling and analytical theory produces results that agree very well with experiment, with a Brillouin gain of $g_{B0} = 0.043\ \mathrm{m^{-1}W^{-1}}$ in the PCF studied. How these CFWs are excited by intervortex interference remains to be fully understood, although it is clear that both electrostriction and gradient-pressure will play a role.

Intervortex photon-phonon interactions in chiral PCF offer new opportunities for controlling optical signals in advanced multimode photonic systems. For example, once a CFW is excited in the presence of forward-propagating "writing" signals, it can be used for non-reciprocal backward conversion of vortex modes at a different wavelength that can be 51.5 nm away from the "writing" signals, with 42 dB suppression of conversion in the forward direction. This is the first observation of two intermodal SBS conversions that occur at different wavelengths with such large separation, while both processes use only one acoustic phonon wave. The conversion efficiencies can be increased by simply raising the power in the writing beams, and by optimising the design of the core structure and reducing the fibre loss.

The results reported open up new perspectives on stimulated Brillouin scattering as well as providing a route to novel applications in fields such as advanced multimode fibre telecommunications, optoacoustic signal processing[34–38] and multi-dimensional vectorial optical fibre sensing[20]. The theory of optoacoustic coupling between hBMs provides new insight into recent work on circular-polarization-sensitive SBS[39,40].

# Methods

## 1. Measurement of frequency conversion of vortex states

A custom-built setup was used to investigate intervortex frequency conversion by FSBS (Fig. 1). In the writing step, the output from a CW laser at 1550 nm was split into pump and Stokes signals. The pump was frequency upshifted by 10 GHz in a single sideband modulator (SSBM) and the Stokes by $(10 - \Omega)$ GHz. Brillouin gain spectra were measured by sweeping $\Omega$ over the peak of the Brillouin gain. Both pump and Stokes were boosted to few-W levels in erbium-doped fibre amplifiers (EDFAs). Fibre polarisation controllers (FPC) and vortex generation modules (VGM) were optionally used to synthesize vortex beams with different topological charges. The VGMs consisted of a sequence of polariser, λ/4 plate, Q-plate, and λ/4 plate. A circular-polarizing beam splitter (CPBS) was used to set orthogonal circular pump and Stokes polarisation states. At the output of the chiral PCF, a λ/4 plate and a PBS were adjusted to transmit the pump and reflect the Stokes signals. For the measurement of intervortex FSBS between [0,+1] and [±1,−1] modes, the amplification of Stokes was measured by power meter (PM), while for the case of [+1,+1] to $[-1,-1]$ scattering, a double-lock-in detection scheme[18] was used. In this scheme, two low-frequency sinusoidal signals ($f_a$ = 20 kHz and $f_b$ = 15 kHz), generated in a lock-in amplifier (LIA), were superimposed separately on to the pump and Stokes signals using intensity modulators (IM). The LIA measures the Brillouin response from the Stokes signal at the output, with a lock-in frequency of 5 kHz ($= (20 - 15)$ kHz). This double-lock-in technique allows not only measurement of the weak FSBS signal, but also rapid location of the Brillouin peak before measurement of the gain spectrum.

In the read-out step, the output from a frequency tuneable CW laser was launched into the fibre in the backward direction, keeping the forward pump and Stokes switched on with their frequency difference set to the peak of the Brillouin gain. A VGM was optionally used to create circularly-polarized vortex states. The backward pump/Stokes was frequency-tuned until conversion to a frequency-shifted Stokes/pump was observed, due to the presence of the already-written acoustic wave. The power and spectrum of backward pump and Stokes signals were measured during read-out by a power meter and an optical spectrum analyser (OSA). Again, the lock-in detection scheme was optionally used to measure backward intervortex conversion if the signal response was too weak, as was the case $[+1,+1]$ to $[-1,-1]$ conversion. A near-field scanning Brillouin analyser (NBA)[28] was used to measure the near-field distribution and topological charge of the modes of the newly-generated signals. See Methods section 2 for more details on NBA.

## 2. Measurement of mode profiles

As Fresnel reflections and Rayleigh scattering made direct measurement of the near-field mode profiles of the read-out signals difficult, we employed a near-field scanning Brillouin analyser (NBA). This consists of an objective lens, a fibre raster scanning stage controlled by a computer with closed-loop feedback, a narrow-band filter, and signal detection equipment (e.g., OSA or PM). Light from the fibre was collected pixel by pixel by the fibre raster scanning stage. The signal was then filtered to remove unwanted light and detected and analysed. The NBA was also used to measure the topological charge, requiring only the addition of a cylindrical lens in front of it to create the patterns characteristic of orbital angular momentum.

**Acknowledgments:** The authors thank Michael H. Frosz for providing the twisted PCF and Andreas Geilen, Steven Becker, Dr. Hagai Diamandi and Dr. Yosef London for helpful discussions concerning the experiments.

**Funding:** This research was supported by the Max-Planck-Gesellschaft through an independent Max-Planck-Research Group.

**Author contributions:**

Conceptualisation: X.Z., P.R., B.S

Experiment and numerical modelling: X.Z.

Theory: P.R.

Analysis: X.Z., P.R., B.S

Funding acquisition: P.R., B.S

Project administration: P.R., B.S

Supervision: P.R., B.S

Writing – original draft: X.Z., P.R., B.S

Writing – review & editing: X.Z., P.R., B.S

**Competing interests:** Authors declare no competing interests.

**Supplementary Materials**

Sec.1 and 2

Figs. S1 to S3

References (1-3)

# Supplementary Materials for

## Frequency conversion of vortex states by chiral forward Brillouin scattering in twisted photonic crystal fibre

Xinglin Zeng[1*], Philip St.J. Russell[1] and Birgit Stiller[1,2*]

*Correspondence to: xinglin.zeng@mpl.mpg.de, birgit.stiller@mpl.mpg.de

**This PDF file includes:**

Supplementary Text Sec.1 and 2

Figs. S1 to S3

References (1-3)

# 1. Theory of intervortex FSBS

## 1.1 LCP to RCP plane-wave coupling by linear birefringence wave

Two orthogonal circularly polarised plane waves of different frequency can be coupled together by a weak acoustic wave that induces a rotating linear birefringence. The dielectric constant of such a system can be written:

$$[\varepsilon] = [\varepsilon_{0c}] + [\varepsilon_{ph}] = \begin{bmatrix} \varepsilon_0 & i\varepsilon_c/2 \\ -i\varepsilon_c/2 & \varepsilon_0 \end{bmatrix} + \begin{bmatrix} \cos\zeta & -\sin\zeta \\ \sin\zeta & \cos\zeta \end{bmatrix} \begin{bmatrix} \varepsilon_1/2 & 0 \\ 0 & -\varepsilon_1/2 \end{bmatrix} \tag{1}$$

where $\zeta = Kz - \Omega t$, $\varepsilon_0$ is the average dielectric constant, $\varepsilon_c$ causes circular birefringence and $\varepsilon_1$ a weak linear birefringence, and $K$ is the acoustic wavevector (acoustic phase velocity $\Omega/K = (\omega_+ - \omega_-)/K$). For $\varepsilon_1 = 0$, Maxwell's equations yield LC and RC polarised fields with propagation constants $\beta_\pm = n_\pm \omega/c = (\omega/c)\sqrt{\varepsilon_0 \pm \varepsilon_c/2}$, where $n_i$ is the refractive index. The resulting circular birefringence helps maintain circular polarisation states against perturbations. Assuming a transverse field in the form:

$$\mathbf{e} = a_+\mathbf{e}_+ + a_-\mathbf{e}_- = a_+(z)\frac{1}{\sqrt{2}}\begin{pmatrix} 1 \\ i \end{pmatrix} e^{i(\beta_+ z - \omega_+ t)} + a_-(z)\frac{1}{\sqrt{2}}\begin{pmatrix} 1 \\ -i \end{pmatrix} e^{i(\beta_- z - \omega_- t)} \tag{2}$$

where $a_+$ and $a_-$ are slowly varying pump and Stokes amplitudes, coupling between the two waves will be proportional to $\langle \mathbf{e}_+^\dagger | \varepsilon_{ph} | \mathbf{e}_- \rangle$, resulting in the equations:

$$\frac{\partial a_+}{\partial z} = i\kappa_+ a_- e^{i\vartheta z}, \qquad \frac{\partial a_-}{\partial z} = i\kappa_- a_+ e^{-i\vartheta z} \tag{3}$$

with coupling constants $\kappa_i = \omega_i \varepsilon_1/(2n_i c)$ and dephasing rate $\vartheta = (\beta_- - \beta_+ \pm K)$, the sign being chosen to minimize $|\vartheta|$. Coupling between LCP and RCP modes can only occur if there is some linear birefringence.

## 1.2 Optoacoustic coupling between hBMs

We now extend this analysis to coupling between circularly polarised helical Bloch modes (hBMs) carrying optical vortices and guided in an *N*-fold rotationally symmetric chiral PCF. In the Cartesian laboratory frame, the transverse field of such a hBM can be written as a superposition of Bloch harmonics, each of which carries an optical vortex[1]:

$$\begin{aligned} \mathbf{e}_k &= a_k(z)\mathbf{e}_{Bk} = a_k(z)\begin{pmatrix} e_{Bkx} \\ e_{Bky} \end{pmatrix} \\ &= a_k(z)\begin{pmatrix} 1 \\ s_k i \end{pmatrix}\frac{1}{\sqrt{2}}\sum_{m=-\infty}^{\infty}\sum_{q=1}^{\infty} b_{q,\ell_k^{(m)}} J_{\ell_k^{(m)}}\left(\frac{u_{q,\ell_k^{(m)}}\rho}{\rho_0}\right) e^{i\left(\beta_k^{(m)} z + \ell_k^{(m)}\phi - i\omega_k t\right)} \\ &= a_k(z)\begin{pmatrix} 1 \\ s_k i \end{pmatrix}\frac{1}{\sqrt{2}} B_k(\rho, \phi + \alpha z) e^{i(\beta_k z + \ell_k \phi - \omega_k t)} \end{aligned} \tag{4}$$

where $a_k(z)$ is a slowly-varying amplitude, $B_k(\rho, \phi + \alpha z)$ is the azimuthally periodic field distribution of the *k*-th hBM ($k = P, S$), $(\rho, \phi, z)$ are cylindrical coordinates, $\omega_k$ the angular frequency, $s_k = \pm 1$ the spin, and the *m*-th orbital harmonic has propagation constant $\beta_k^{(m)} = \beta_k + mN\alpha$, topological charge $\ell_k^{(m)} = \ell_k + mN$ and amplitude $b_{q,\ell_k^{(m)}}$ (to simplify the notation we have set $\beta_k^{(0)} = \beta_k$ and $\ell_k^{(0)} = \ell_k$). The radial distribution is given by the summation $q$, where $u_{q,\ell}$ is the $q$-th zero of the $J_\ell$ Bessel function, and the effective edge of the photonic crystal region, where the fields vanish, is set at $\rho = \rho_0$ (this should be large enough to accommodate the guided modes). Note that each harmonic in $B_k$ contains the factor $e^{imN(\alpha z+\phi)}$, i.e., so that the hBM field pattern rotates with $z$ at a rate $\partial\phi/\partial z = -\alpha$, as required of the chiral fibre.

In the absence of any perturbation ($a_P$ and $a_S$ constant), the expression in Eq. (4) is itself a solution of Maxwell's equations. As in Eq. (3) of main text, the coupling rate from mode $k$ to mode $l$ will be proportional to $\langle \mathbf{e}_{Bl}^{\dagger}|\varepsilon_{ph}|\mathbf{e}_{Bk}\rangle$ where † denotes the Hermitian conjugate and this time the perturbation $[\varepsilon_{ph}]$ has a transverse profile (caused by the guided acoustic wave) and $\zeta = Kz + \ell_{ph}\phi - \Omega t$, where $\ell_{ph}$ is the acoustic topological charge and $\varepsilon_1$ is the induced anisotropic change in dielectric constant that yields linear birefringence. Note that it is important to ensure that $K$ corresponds to the acoustic harmonic that carries topological charge $\ell_{ph}$, since adjacent harmonics differ in wavevector by the twist rate $N\alpha$. Coupling between orthogonal circularly polarised vortex modes can only occur for $|\varepsilon_1| > 0$, and the rotation matrix in Eq. (1) ensures that the pattern of linear birefringence rotates with position and time, resembling a rotating single-spiral corkscrew. A chiral acoustic wave with this topology can be produced by a superposition of two, π/2 out-of-phase, orthogonal flexural waves differing in frequency by Ω (see next section).

Evaluating $\langle \mathbf{e}_{l}^{\dagger}|\varepsilon_{ph}|\mathbf{e}_{k}\rangle$ in a coupled-mode description, assuming slow-varying power-normalised amplitudes, a separately excited acoustic wave (i.e., no SBS), conservation of topological charge, i.e., $\ell_{ph} = \ell_S - \ell_P$, and collecting terms with the slowest rates of phase progression, we obtain:

$$\frac{\partial a_P}{\partial z} + \frac{\alpha_P}{2}a_P = i\kappa_P a_S e^{i\vartheta z}, \qquad \frac{\partial a_S}{\partial z} + \frac{\alpha_S}{2}a_S = i\kappa_S a_P e^{-i\vartheta z}. \tag{5}$$

The quantity $\vartheta = \beta_S(\omega_S) - \beta_P(\omega_P) - K(\Omega)$ is the dephasing rate, $\beta_i(\omega_i) = \omega_i n_i/c$ is the propagation constant and $n_i$ the phase index of mode $i$, and $\kappa_i = \omega_i Q/c$, where $Q$ is a dimensionless parameter that is proportional to the amplitude overlap integral and the induced linear birefringence. Note that the optoacoustic overlap integral is calculated using the entire field of each helical Bloch mode, i.e., the sum over all harmonics.

In analytically solving Eq. (5) it is often convenient to make the substitution $a_P = b_P e^{i\vartheta z}$, resulting in:

$$\frac{\partial b_P}{\partial z} + \left(\frac{\alpha_P}{2} + i\vartheta\right)b_P = i\kappa_P a_S, \qquad \frac{\partial a_S}{\partial z} + \frac{\alpha_S}{2}a_S = i\kappa_S b_P. \tag{6}$$

### 1.3 Equations of FSBS: Writing process

The coupled SBS equations for power-normalised field amplitudes $a_i$ can be written in the form:

$$\frac{\partial a_P}{\partial z} + \frac{\alpha_P}{2}a_P = i\frac{\omega_P}{c}\sqrt{Q_B}a_S a_{ph}, \qquad \frac{\partial a_S}{\partial z} + \frac{\alpha_S}{2}a_S = i\frac{\omega_S}{c}\sqrt{Q_B}a_P a_{ph}^*,$$
$$\frac{\partial a_{ph}}{\partial z} + \left(\frac{\alpha_{ph}}{2} + i\vartheta\right)a_{ph} = i\frac{\omega_{ph}}{c}\sqrt{Q_B}a_S^* a_P \tag{7}$$

where $\vartheta = \beta_P - \beta_S - \beta_{ph}$ is the dephasing parameter, $c$ is the speed of light in vacuum, and $Q_B$ is a characteristic parameter (with units W$^{-1}$) that depends on the electrostrictive parameters and the optoacoustic overlap. Noting that the power in the $i$-th mode is proportional to $p_i = |a_i|^2$, power conservation can be checked by multiplying the first equation by $a_P^*$, adding the result to its complex conjugate, and doing the same for the other three equations. Adding all the equations together we obtain:

$$\frac{\partial}{\partial z}\left(|a_P|^2+|a_S|^2+\left|a_{ph}\right|^2\right)+\alpha_P|a_P|^2+\alpha_S|a_S|^2+\alpha_{ph}\left|a_{ph}\right|^2 = c^{-1}\sqrt{Q_B}(\omega_P-\omega_S-\omega_{ph})ia_P^*a_Sa_{ph}+\mathrm{c.c.}=0 \tag{8}$$

since $\omega_P-\omega_S-\omega_{ph}=0$, showing that power is conserved in the absence of loss.

With the good approximation (commonly used in SBS) that $|\partial/\partial z|\ll\left|\alpha_{ph}/2+i\vartheta\right|$, Eq. (7) can be recast in the form:

$$\begin{aligned}\frac{\partial a_P}{\partial z}+\frac{\alpha_P}{2}a_P&=-\frac{\omega_P\omega_{ph}Q_B/c^2}{\alpha_{ph}/2+i\vartheta}(a_Pa_S^*)a_S=-\frac{2\omega_P\omega_{ph}Q_B(1-i2\vartheta/\alpha_{ph})}{c^2\alpha_{ph}(1+4\vartheta^2/\alpha_{ph}^2)}a_P|a_S|^2\\ \frac{\partial a_S}{\partial z}+\frac{\alpha_S}{2}a_S&=\frac{\omega_S\omega_{ph}Q_B/c^2}{\alpha_{ph}/2-i\vartheta}(a_Sa_P^*)a_P=\frac{2\omega_S\omega_{ph}Q_B(1+i2\vartheta/\alpha_{ph})}{c^2\alpha_{ph}(1+4\vartheta^2/\alpha_{ph}^2)}a_S|a_P|^2.\end{aligned} \tag{9}$$

Multiplying the first by $a_P^*$ and the second by $a_S^*$ and adding each equation to its complex conjugate, we obtain the coupled power equations:

$$\begin{aligned}\frac{\partial p_P}{\partial z}+\alpha_Pp_P&=-\frac{4\omega_P\omega_{ph}Q_B/c^2}{\alpha_{ph}(1+4\vartheta^2/\alpha_{ph}^2)}p_Pp_S\\ \frac{\partial p_S}{\partial z}+\alpha_Sp_S&=\frac{4\omega_S\omega_{ph}Q_B/c^2}{\alpha_{ph}(1+4\vartheta^2/\alpha_{ph}^2)}p_Pp_S.\end{aligned} \tag{10}$$

Although, because energy is lost to phonons, these equations no longer exactly conserve power for $\alpha_P=\alpha_S=0$, they do (as required) conserve photon flux:

$$\frac{\partial}{\partial z}\left(\frac{p_P}{\hbar\omega_P}+\frac{p_S}{\hbar\omega_S}\right)=0. \tag{11}$$

Equations (10) can be cast in a simplified form by noting that $\omega_P\cong\omega_S\cong(\omega_P+\omega_S)/2=\bar{\omega}$:

$$\frac{\partial p_P}{\partial z}+\alpha_Pp_P=-g_Bp_Pp_S=-\frac{g_{B0}}{1+4\vartheta^2/\alpha_{ph}^2}p_Pp_S,\quad \frac{\partial p_S}{\partial z}+\alpha_Sp_S=g_Bp_Pp_S \tag{12}$$

Where

$$g_B=\frac{g_{B0}}{1+4\vartheta^2/\alpha_{ph}^2}=\frac{4\bar{\omega}\omega_{ph}Q_B/(c^2\alpha_{ph})}{1+4\vartheta^2/\alpha_{ph}^2}\,\mathrm{m^{-1}W^{-1}}. \tag{13}$$

Equations (12) have an instructive analytical solution if the loss in both pump and Stokes modes is identical, i.e., $\alpha_P=\alpha_S=\alpha_0$:

$$p_P(z)=P_0\frac{p_{P0}e^{-\alpha_0 z}}{p_{P0}+(P_0-p_{P0})e^{g_BP_0L_{\mathrm{eff}}(z)}},\quad p_S(z)=P_0e^{-\alpha_0 z}-p_P(z) \tag{14}$$

where $p_P(0)=p_{P0}$, $p_S(0)=p_{S0}$, $P_0=p_{S0}+p_{P0}$ is the total input power, and $L_{\mathrm{eff}}(z)=(1-e^{-\alpha_0 z})/\alpha_0$.

### 1.4 Phase of excited acoustic wave

To analyse backward coupling, which is essentially a linear scattering process, we need to know the strength and phase of the CFW at every point along the fibre. This requires solution of the SBS amplitude equations (Eq. (9)). Setting $a_i=q_i(z)e^{i\psi_i(z)}$ where $q_i$ and $\psi_i$ are real-valued, substituting into Eq. (9) and extracting the imaginary parts, we obtain:

$$\frac{\partial\psi_P}{\partial z} = \vartheta\frac{g_B}{\alpha_{ph}}q_S^2\,, \qquad \frac{\partial\psi_S}{\partial z} = \vartheta\frac{g_B}{\alpha_{ph}}q_P^2, \tag{15}$$

which shows that in general the relative phase of pump and Stokes varies in a non-trivial manner along the fibre. As a result, the phase of the excited acoustic wave is not constant, which will affect the conversion efficiency of backward conversion. However, in the experiment the dephasing is zero in the writing process, so that the pump and Stokes phases are constant.

### 1.5 Electrostrictive driving term

Given the complexity of the PCF structure, standard approaches to electrostriction are not valid and numerical techniques must be used[2]. We can nevertheless make some general observations, starting with an electric field created by a superposition of two vortex modes with different topological charge, spin and frequency:

$$\mathbf{E} = a_1\begin{pmatrix}1\\ i\end{pmatrix}\frac{1}{\sqrt{2}}e^{iq_1} + a_2\begin{pmatrix}1\\ -i\end{pmatrix}\frac{1}{\sqrt{2}}e^{iq_2} \tag{16}$$

where $q_i = \beta_i z + \ell_i\phi - \omega_i t$. Collecting and combining the two vector components we obtain:

$$\mathbf{E} = e^{i\bar{q}}\frac{1}{\sqrt{2}}\begin{pmatrix} a_1 e^{i\delta q/2} + a_2 e^{-i\delta q/2} \\ i(a_1 e^{i\delta q/2} - a_2 e^{-i\delta q/2}) \end{pmatrix} \tag{17}$$

where $\bar{q} = (q_1 + q_2)/2$ and $\delta q = (q_1 - q_2)$. Taking the case when $a_1 = a_2 = a$ for illustrative purposes, Eq. (17) yields:

$$\mathbf{E} = e^{i\bar{q}}a\sqrt{2}\begin{pmatrix}\cos\delta q/2\\ -\sin\delta q/2\end{pmatrix} \tag{18}$$

which is a fast-oscillating linearly polarised field whose average direction rotates slowly with time, azimuthal angle and distance. Electrostriction cannot follow the fast oscillation $\bar{q}$, but it is able to respond to the much slower (up to ~2 GHz) rotation $\delta q$, which we propose induces an anisotropic rotating strain field that when phase-matched can drive a CFW, and cause an anisotropic change in dielectric constant similar to Eq. (1). Understanding this mechanism fully will require further study.

### 1.6 Intervortex conversion by written CFW: Read-out process

The read-out process can be analysed by extracting the coupling term in Eq. (9) and writing coupled equations for the read-out amplitudes $b_i$:

$$\begin{aligned} \frac{\partial b_P}{\partial z} + \frac{\alpha_{Pr}}{2}b_P &= -\left(\frac{2\bar{\omega}_r\omega_{ph}Q_B}{\alpha_{ph}c^2}a_P a_S^*\right)b_S \mathrm{e}^{i\vartheta_r z} \cong -\left(\frac{g_{B0}}{2}a_P a_S^*\right)b_S\mathrm{e}^{i\vartheta_r z} \\ \frac{\partial b_S}{\partial z} + \frac{\alpha_{Sr}}{2}b_S &= \left(\frac{2\bar{\omega}_r\omega_{ph}Q_B}{\alpha_{ph}c^2}a_S a_P^*\right)b_P\mathrm{e}^{-i\vartheta_r z} \cong \left(\frac{g_{B0}}{2}a_S a_P^*\right)b_P\mathrm{e}^{-i\vartheta_r z} \end{aligned} \tag{19}$$

where $\vartheta_r = \beta_{Sr} - \beta_{Pr} \pm K$, the sign being chosen to minimize $|\vartheta_r|$, and $K$ is set in the writing step, as is the acoustic frequency $\Omega$. In deriving Eq. (19) we have assumed that the writing process is perfectly phase-matched, i.e., $\vartheta = 0$, and made the approximation $\omega_{Pr} \cong \omega_{Sr} \cong (\omega_{Pr} + \omega_{Sr})/2 = \bar{\omega}_r$. The terms in the large brackets are proportional to the strength of the already-written acoustic wave and play the same role as $i\kappa_P$ and $i\kappa_S$ in Eq. (3) of the main text; note that the amplitude gain is one half of the power gain $g_{B0}$. Since the acoustic frequency and wavevector are fixed by the writing step and the optical dispersion is much weaker than the acoustic, the dephasing $\vartheta_r$ changes much more slowly with the optical frequency than in the writing step.

## 2. Calculation of acoustic modes and optoacoustic overlaps

We used finite element modelling (FEM) to calculate the optoacoustic overlap integrals in untwisted PCF. Usually, Brillouin scattering originates from two main physical effects: electrostriction and radiation pressure. The large air-filling fraction and small core of twisted PCF requires more comprehensive calculations to clarify the contributions from both physical effects. The following formulae were used in the calculations[2,3]:

$$Q_{\mathrm{ES}} = \left| \iint \frac{\mathbf{E}_{\mathrm{P}}^{*} \cdot \delta \boldsymbol{\varepsilon}_{\mathrm{ES}}^{*} \cdot \mathbf{E}_{\mathrm{S}}}{\max\left(|\mathbf{u}|\right) N_{\mathrm{P}} N_{\mathrm{S}}} dA \right|^2$$

$$Q_{\mathrm{RP}} = \left| \int \frac{\mathbf{u}^{*} \cdot \hat{n}\left(\delta\varepsilon_{\mathrm{RP}} \mathbf{E}_{\mathrm{P},\parallel}^{*} \cdot \mathbf{E}_{\mathrm{S},\parallel} - \delta\varepsilon_{\mathrm{RP}}^{-1} \mathbf{D}_{\mathrm{P},\perp}^{*} \cdot \mathbf{D}_{\mathrm{S},\perp}\right)}{\max\left(|\mathbf{u}|\right) N_{\mathrm{P}} N_{\mathrm{S}}} dl \right|^2 \qquad (20)$$

where $\mathbf{E}_{j,\parallel}^{*}$ and $\mathbf{D}_{j,\perp}^{*}$ are the tangential electric field and normal electric displacement field at the interfaces between the hollow channels and the glass; $j = P$ denotes the pump and $j = S$ the Stokes, and $\mathbf{u}^{*} \cdot \hat{n}$ is the surface normal component of the displacement vector $\mathbf{u}^{*}$. Changes in permittivity due to radiation pressure are given by $\delta\varepsilon_{\mathrm{RP}} = \varepsilon_1 - \varepsilon_2$ and $\delta\varepsilon_{\mathrm{RP}}^{-1} = 1/\varepsilon_1 - 1/\varepsilon_2$, where $\varepsilon_1$ and $\varepsilon_2$ are the permittivities of the glass and the hollow channels. The permittivity change due to electrostriction is $\delta\boldsymbol{\varepsilon}_{\mathrm{ES}}^{*} = -\varepsilon_0 n^4 \mathbf{p} \colon \mathbf{S}$, where $n$ is the material refractive index, $\mathbf{p}$ the photoelastic tensor, and $\mathbf{S} = \nabla_s \mathbf{u}$ is the symmetric strain caused by the acoustic waves, all spatially dependent quantities. $N_j = \left(2Re\left(\int \mathbf{E}_{\mathrm{j}} \times \mathbf{H}_{\mathrm{j}}^{*} \cdot \hat{z} dA\right)\right)^{1/2}$ is the optical power. In our case, $Q_{\mathrm{RP}} \ll Q_{\mathrm{ES}}$. Fig. S1 shows the normalised overlap coefficients ($Q_{\mathrm{RP}}$ in red, $Q_{\mathrm{ES}}$ in blue) of all the acoustic modes that satisfy phase-matching when the pump/Stokes is [0,+1]/[−1,−1] (Fig. S1(a)) and [+1,+1]/[−1,−1] (Fig. S1(b)). The acoustic wavevector is first set to $\Delta\beta = 2\pi(n_P - n_S)/\lambda_P$, and the frequencies of the acoustic modes that share this wavevector are calculated. Fig. S2 shows total displacements of the three acoustic modes having high optoacoustic overlap in Fig. S1(a). The acoustic mode with cut-off frequency 95.8 MHz is mostly confined in the core and closest to the phase-matching condition (Fig. 3(a) in the main text) and dominates FSBS. However, the other two do play a role, creating a shoulder on the high frequency side of the gain spectrum (Fig. 3(d) in the main text). Fig. S3 shows the strain energy densities, transverse and axial displacements of six acoustic modes having high optoacoustic overlaps in Fig. S1(b). Similarly, at 1.292 GHz one acoustic mode dominates the FSBS response while neighbouring ones enlarge the bandwidth.

Note that those calculations are only valid for uniaxial flexural acoustic modes in untwisted PCF. Chiral flexural acoustic modes can then be formed by the superposition of two orthogonal, π/2-out-of-phase uniaxial flexural modes with different frequencies. In the chiral PCF the acoustic modes are expected to display some topological birefringence, which although not discussed here will have some small effect on the phase-matching conditions.

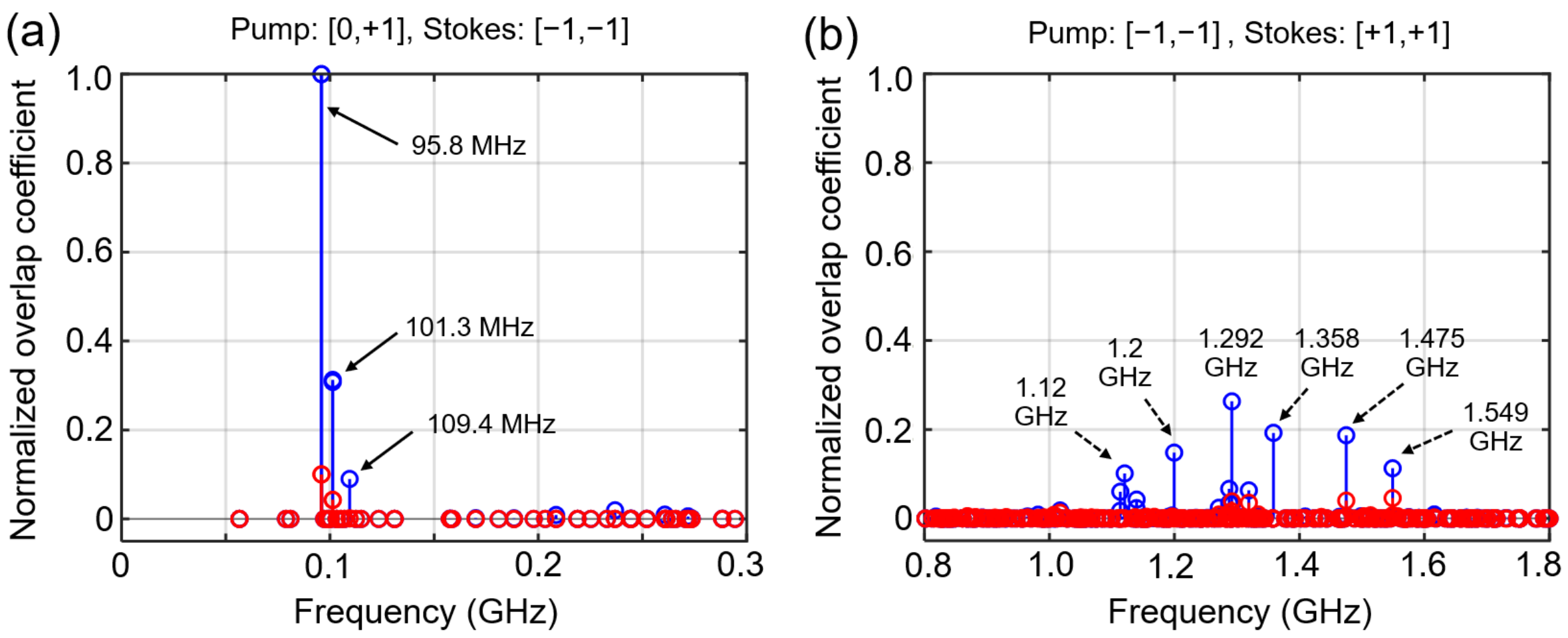

**Fig. S1.** (a) Numerically calculated overlap coefficients ($Q_{RP}$ in red, $Q_{ES}$ in blue) for all the acoustic modes that satisfy phase-matching in the range 0 to 0.3 GHz, for [0,+1] pump and [−1,−1] Stokes modes. (b) Same as (a) but for [−1,−1] pump and [−1,−1] Stokes modes, and acoustic modes in the range 0.8 to 1.8 GHz.

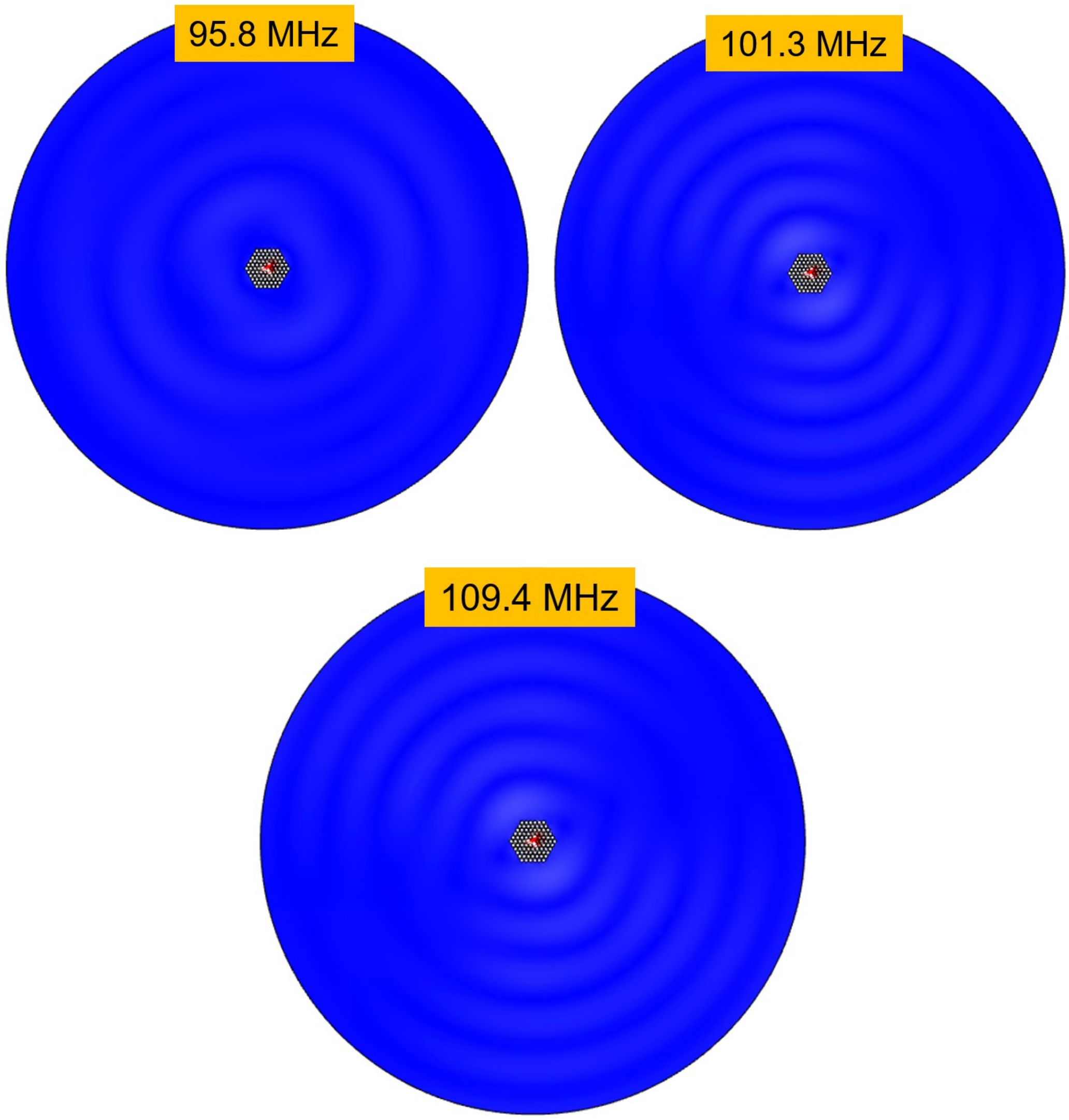

**Fig. S2.** Total displacements of three acoustic mode at 95.8 MHz, 101.3 MHz and 109.4 MHz. All of them have similar displacement distributions in the core region, but the one at 98.5 MHz is the most tightly confined (35.8%) to the core and has highest overlap with the optical modes. The other two modes are less tightly confined (25.4% and 13.6%) to the core and are distributed over the whole fibre cross-section.

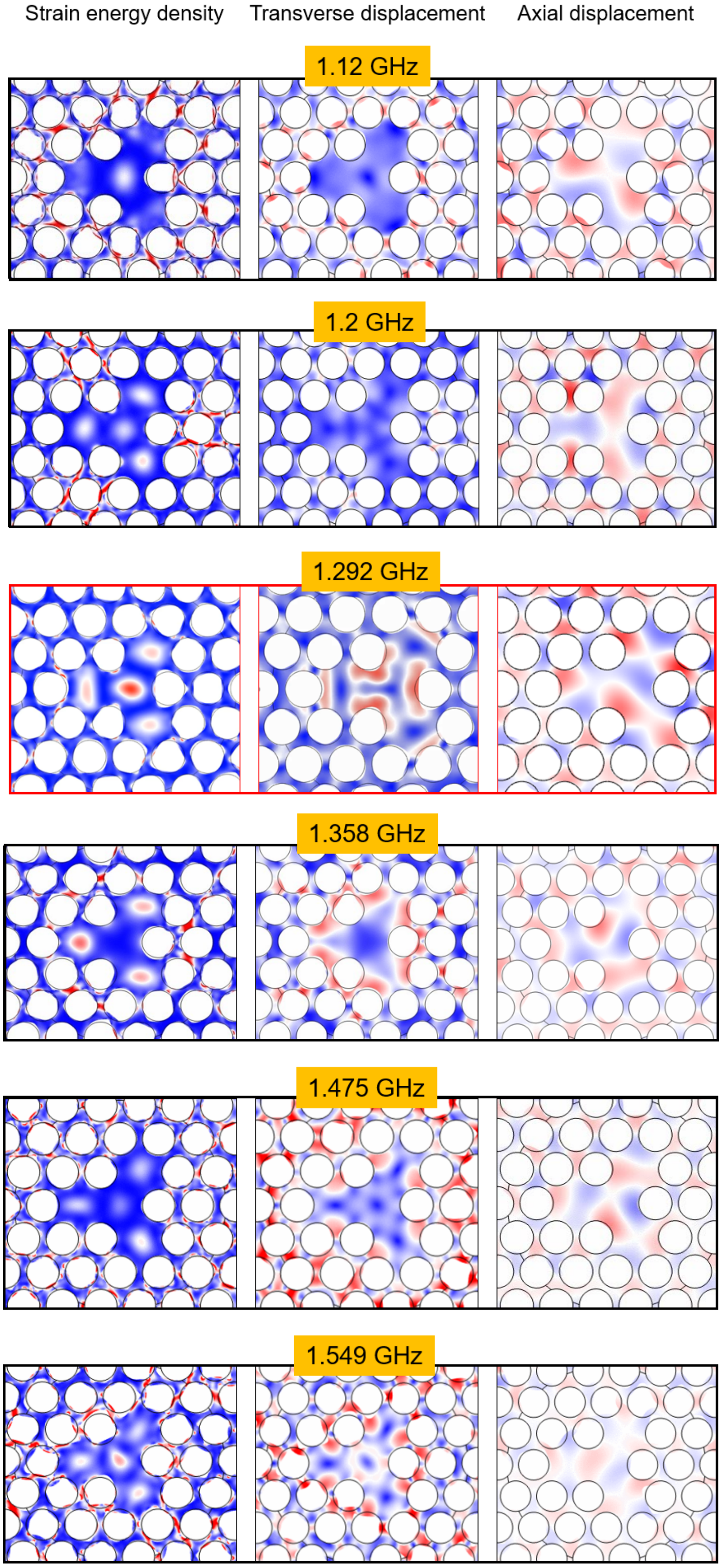

**Fig. S3.** The strain energy density, transverse and axial displacements of all six labelled CFWs in Fig. S1(b). They all have relatively high optoacoustic overlaps with the [−1,−1] pump and [+1,+1] Stokes modes. The acoustic mode at 1.292 GHz has the highest overlap and dominates FSBS.